\documentclass[fleqn,usenatbib]{mnras}

\usepackage{newtxtext,newtxmath}

\usepackage[T1]{fontenc}

\DeclareRobustCommand{\VAN}[3]{#2}
\let\VANthebibliography\thebibliography
\def\thebibliography{\DeclareRobustCommand{\VAN}[3]{##3}\VANthebibliography}

\usepackage{graphicx}	
\usepackage{amsmath}	
\usepackage{amssymb}	
\usepackage{subfig}

\title[Inclination damping on Callisto]{Inclination damping on Callisto}

\author[B. G. Downey et al.]{
Brynna G. Downey,$^{1}$\thanks{E-mail: bgdowney@ucsc.edu (BGD)}
F. Nimmo,$^{1}$
Isamu Matsuyama$^{2}$
\\
$^{1}$Department of Earth and Planetary Sciences, University of California, Santa Cruz, CA, 95064, USA\\
$^{2}$Lunar and Planetary Laboratory, University of Arizona, Tucson, AZ 85719, USA\\
}

\date{Accepted XXX. Received YYY; in original form ZZZ}

\pubyear{2015}

\begin{document}
\label{firstpage}
\pagerange{\pageref{firstpage}--\pageref{lastpage}}
\maketitle

\begin{abstract}
\label{section:abstract}
Callisto is thought to possess a subsurface ocean, which will dissipate energy due to obliquity tides. This dissipation should have damped any primordial inclination within 1~Gyr - and yet Callisto retains a present-day inclination. We argue that Callisto's inclination and eccentricity were both excited in the relatively recent past ($\sim 0.3$~Gyr). This excitation occurred as Callisto migrated outwards according to the ``resonance-locking'' model and passed through a 2:1 mean-motion resonance with Ganymede. Ganymede's orbital elements were likewise excited by the same event. To explain the present-day orbital elements we deduce a solid-body tidal $k_2/Q \approx 0.05$ for Callisto, and a significantly lower value for Ganymede.

\end{abstract}




\section{Introduction}
\label{section:introduction}

The thermal and orbital evolution of satellites is governed by energy dissipated by tides in both the planet and the satellite \citep[e.g.,][]{goldreich1966q}.

Tides raised on the planet by the satellite generally lower the planet's spin rate and transfer angular momentum to the satellite, increasing its semi-major axis (the orbital distance). The exceptions to this rule are retrograde satellites, such as Triton, and satellites inside the synchronous rotation point, such as Phobos. In these cases, the satellite spirals in towards the planet. Tides raised on the planet will also raise the satellite's eccentricity (how far the elliptical orbit is from being circular) and lower its inclination (the angle between the orbital plane and the Laplace plane). The Laplace plane is the mean orbital plane, so the orbit normal precesses around the Laplace plane normal.

Tides raised on a synchronously-rotating satellite by the planet will lower either the inclination or the eccentricity depending on whether obliquity (the angle between the orbit normal and the spin pole) or eccentricity is at the root of the synchronous rotation anomaly. A satellite's obliquity and inclination are related via Cassini states \citep{ward1975past}, so obliquity tides lower inclination, which lowers obliquity \citep[e.g.,][]{chyba1989tidal}. 

The amount that the inclination and eccentricity decrease by depends on how easily the satellite deforms due to the planet's gravitational pull and how much friction its interior experiences in trying to realign the tidal bulge to the line connecting the centres of mass. Rates of change for inclination and eccentricity and their corresponding tidal heating are most frequently computed under the assumption that satellites are completely solid, viscoelastic bodies \citep{ross1986tidal}. 

There is growing evidence, however, that many satellites in our solar system are not purely solid bodies and may have subsurface oceans \citep{nimmo2016ocean}. For example, detections of an induced magnetic field by the Galileo magnetometer during Callisto flybys \citep{zimmer2000subsurface} and the measurement of Titan's obliquity \citep{bills2011rotational,baland2011titan} strongly suggest that Callisto and Titan have subsurface oceans. Although \cite{hartkorn2017induction} propose that the magnetic field signal detected at Callisto could be accounted for by induction in its ionosphere, in this work we will assume that Callisto has a subsurface ocean \citep{zimmer2000subsurface}. Satellites without the advantage of extensive flybys, such as Oberon and other outer satellites might still have subsurface oceans according to models that emphasize the role of salts in reducing the melting temperature of ice \citep{hussmann2006subsurface}.

As discussed in more detail below, dissipation in subsurface oceans can be substantial and, crucially, tends to damp orbital inclination as or even more rapidly than the solid body does. Thus, for a body like Callisto with a subsurface ocean, the survival of a non-zero present-day inclination presents a puzzle. The bulk of this manuscript investigates how such a non-zero inclination could be maintained.

In the remainder of this section we lay out the basics of inclination and eccentricity damping. In Section~\ref{section:lid}, we show that Callisto's inclination damping time-scale is expected to be short compared to the age of the solar system for likely parameter values. In Section~\ref{section:fuller}, we show that in a frequency-dependent $Q$ of Jupiter scenario, one or more mean-motion resonance crossings could increase Callisto's inclination to current levels. In Section~\ref{section:discussion}, we show that Callisto's eccentricity and Ganymede's orbital elements can also be explained by these resonance crossings. We conclude by suggesting further work and making predictions that can be tested with future spacecraft missions.

\subsection{Ocean tidal dissipation and inclination damping}

With an increased number of suspected subsurface oceans in our solar system comes the question of how energy is dissipated in the non-solid body. Here we review a few important contributions to this subject. \cite{sagan1982tide} and \cite{sohl1995tidal} calculated the dissipation of eccentricity tides in Titan's presumed methane surface ocean to determine the lifetime of its eccentricity. \cite{tyler2008strong, tyler2009ocean, tyler2011tidal} emphasized the importance of obliquity tides and made the first numerical models of tidal heating in a surface ocean. \cite{chen2014tidal} expanded on this by numerically deriving formulas for eccentricity and obliquity tide dissipation due to bottom drag in a surface ocean. 

Leading up to the model that we use in this work, \citet{matsuyama2014selfgravity} considered the effects of self-gravity and deformation of the solid regions assuming linear drag, and \citet{hay2017numerically} developed a numerical model that takes these effects into account for both linear and bottom drag. \citet{beuthe2016subsurf} provided the first rigorous quantification of the effect of an overlying ice shell by treating the ice shell as a massless membrane. \citet{matsuyama2018subsurface} expanded on this by providing a theoretical treatment that is applicable to elastic shells of arbitrary thickness, and \citet{hay2019subsurface} used this theory to consider dissipation in a subsurface ocean due to bottom drag with a numerical model.

These numerical models for calculating the tidal dissipation rate in an ocean should be compared with the standard rate of tidal dissipation in a solid synchronous satellite, given by
\begin{equation}
\label{diss solid}
    \dot{E}_{\text{solid}} = \frac{3}{2}\frac{k_2}{Q}\frac{\Omega^5R^5}{G}(\sin^2\theta_0+7e^2),
\end{equation}
\noindent
\textcolor{magenta}{\citep{peale1978contribution, peale1979melting, wisdom2004spin}}
where $k_2/Q$ is a measure of how deformable the satellite is, $\Omega$ is the spin frequency, $R$ is the radius, $G$ is the gravitational constant, $\theta_0$ is the satellite's obliquity, and $e$ is its eccentricity. 

A consequence of the factor of 7 in Eq. \ref{diss solid} is that solid-body dissipation damps eccentricity more rapidly than obliquity assuming that both are small and comparable in value. The simple relationship between inclination and eccentricity decay rates to their respective tidal dissipation rates can be found in \citet{chyba1989tidal}. Conversely, ocean obliquity tides are in general much more dissipative than eccentricity tides \citep[e.g.,][]{tyler2011tidal}, so the inclination damps more rapidly than eccentricity.

The key finding is that ocean obliquity tides contribute substantially to satellite inclination damping. Just as \cite{sagan1982tide} and \cite{sohl1995tidal} investigated whether tidal dissipation allowed Titan's eccentricity to have lasted the lifetime of the solar system, we seek to calculate approximately how long it would take obliquity tides in a satellite subsurface ocean to damp inclination.To do this, we define an inclination damping time-scale, $\tau_i$, which is a first order, small-inclination approximation of how long it would take obliquity tide dissipation in a satellite's ocean, $\dot{E}_{\text{obl}}$, to damp its present-day inclination, $i$ \citep{sagan1982tide,chyba1989tidal,sohl1995tidal}:
\begin{equation} \label{taui}
    \tau_i \sim \frac{GMm}{a\dot{E}_{\text{obl}}}i^2,
\end{equation}
\noindent
where $M$ is the mass of the planet, $m$ is the mass of the satellite, and $a$ is the satellite's orbital semi-major axis. In Section~\ref{section:lid} below we provide analytical methods for calculating $\dot{E}_{\text{obl}}$ for subsurface oceans.

\begin{figure}
    \centering
{\includegraphics[width=\columnwidth]{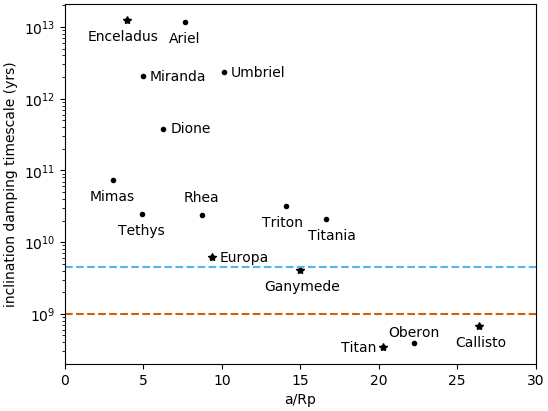}}%
        \caption{Inclination damping time-scales from ocean obliquity tide dissipation rates in \protect\citet{chen2014tidal}. Those icy satellites with evidence of subsurface oceans are denoted with a star \protect\citep{nimmo2016ocean}.  The blue and orange dashed lines are the 4.56~Gyr and 1~Gyr marks, showing that Callisto and Titan are aberrant in having large energy dissipation yet long-lasting inclinations. }%
    \label{fig:tauinc}%
\end{figure}

Fig.~\ref{fig:tauinc} plots the inclination damping time-scales for all icy satellites in our solar system using the ocean dissipation estimates tabulated in \citet{chen2014tidal}. The approximate correlation between damping time-scale and orbital distance appears to be a consequence of the fact that the predicted obliquity relative to the inclination,  and thus the dissipation rate, is itself a strong function of distance. Assuming long-lived oceans, Callisto, Oberon, and Titan all have inclination lifetimes of $\tau_i <1$ Gyr. Any primordial inclination on Callisto, Oberon, and Titan would have been damped out quickly, whereas most other satellites' inclinations could be primordial and have lasted until today. There is no evidence indicating whether Oberon has a subsurface ocean or not, and like the other Uranian satellites, it could have had chaotic orbital evolution \citep{dermott1988dynamics}. Titan may have had a very interesting dynamical history, migrating greatly in semi-major axis \citep{lainey2020resonance} and potentially being influenced by the Jupiter-Saturn Great Inequality \citep{bills2005jupiter}. In this work we focus on Callisto because it presents a somewhat simpler dynamical problem than Titan.

From the above order-of-magnitude calculation, we conclude that Callisto has a non-zero inclination when it should have been damped away by obliquity tides. Either Callisto's physical properties are not well-understood or a dynamical event in the last 1~Gyr increased its inclination. In the remainder of this manuscript we explore both of these possibilities.

\section{Effect of physical properties on ocean dissipation}
\label{section:lid}

\begin{table}
\centering
\caption {Parameters used in Callisto's obliquity tide ocean inclination damping} 
\begin{tabular}{ l l l }
\hline
 Symbol & Parameter & Value \\ 
 \hline
 $G$ & Gravitational constant & 6.674$\times 10^{-11}$ m$^3$ s$^{-2}$ kg$^{-1}$ \\
 $M$ & Mass of Jupiter & 1.898$\times 10^{27}$~kg \\
 $m$ & Mass of Callisto & 1075.9$\times 10^{20}$~kg \\  
 $R$ & Radius & 2410.3 km \\
 $\Omega$ & Spin frequency & 4.36$\times 10^{-6}$~s$^{-1}$ \\
 $\theta_0$ & Obliquity & -0.24$^\circ$ \\
 $g$ & Gravitational acceleration & 1.24 m s$^{-2}$ \\
 $a$ & Semi-major axis & 1882.7$\times 10^6$ m \\
 $e$ & Eccentricity & 0.0074 \\
 $i$ & Inclination & 0.192$^\circ$ \\
 $\rho_b$ & Bulk density & 1830 kg m$^{-3}$\\
 $\rho_o$ & Ocean density & 1000 kg m$^{-3}$ \\
 $\rho_i$ & Ice density & 900 kg m$^{-3}$ \\
$d$ & Ice shell thickness & 150~km \\
 $h$ & Ocean thickness & 30 km \\
 $c_D$ & Bottom drag coefficient & 0.002 \\
$\beta_2$ & Shell pressure forcing coefficient & 0.88 \\
$\upsilon_2$ & Tidal potential forcing coefficient & 1.05 \\
 $\eta$ & Water dynamic viscosity & $10^{-3}$ Pa s \\
 $\mu$ & Ice shear modulus & 3$\times 10^9$ Pa
\end{tabular}
\label{tab:Callisto params}
\end{table}

Callisto's inclination damping time-scale could be the age of the solar system or longer if certain physical properties resulted in reduced dissipation in its putative global ocean. 

From the body of work discussed in the previous section that addresses the ocean obliquity tide dissipation rate, we choose to use the analytical expressions from \citep{hay2019subsurface}. They include the effects of an overlying ice shell, self-gravity, and deformation of the solid regions:
\begin{equation}
\label{Edot_subocean}
\begin{split}
    \dot{E}_{\text{obl}} & =12\pi\rho h\nu_{\text{obl}}\Omega^{2}R^{2}\theta_{0}^{2}\upsilon_{2}^{2}\left(\frac{R}{r_{t}}\right)^{2}\left[1+\left(\frac{20\upsilon_{2}\beta_{2}\nu_{\text{obl}} gh}{\Omega^{3}r_{t}^{4}}\right)^{2}\right]^{-1}\\
\nu_{\text{obl}} & =\frac{\Omega^{3}r_{t}^{4}}{20\sqrt{2}\beta_{2}gh}\left\{ -1+\left[1+\left(\frac{200}{3}0.4c_{D}\beta_{2}\upsilon_{2}\frac{gR^{2}\theta_{0}}{\Omega^{2}r_{t}^{3}}\right)^{2}\right]^{1/2}\right\} ^{1/2}.
\end{split}
\end{equation}
\noindent
Here $\rho$ is the subsurface ocean density, $h$ is the thickness of the ocean, $\nu_{obl}$ is turbulent viscous diffusivity, $\Omega$ is the spin frequency, $R$ is the radius, $\theta_0$ is the obliquity, $g$ is the surface gravity, $r_{t}$ is the ocean top radius, and $c_D$ is the drag coefficient at the bottom of the ocean. Shell pressure forcing is captured by the coefficient $\beta_{2}$, and the perturbation to the forcing tidal potential due to shell pressure forcing, self-gravity, and deformation of the solid regions is captured by the $\upsilon_{2}$ coefficient. These dimensionless coefficients can be computed in terms of pressure and tidal Love numbers \citep[][Eq. 22]{matsuyama2018subsurface}. Assuming a thin surface ocean ($r_{t}\sim R)$ and ignoring self-gravity, deformation of the solid regions, and shell pressure forcing ($\beta_{2}=\upsilon_{2}=1$), Eq.~\ref{Edot_subocean} reduces to the analytical equations in \citet{chen2016tidal} with their factor $\xi_2=1$, as expected. Solutions for a thin surface ocean that take into account the effects of self-gravity and deformation of the solid region can be obtained with the substitutions $r_{t}\rightarrow R$,
$\upsilon_{2}\rightarrow1+k_{2}^{T}-h_{2}^{T}$, and $\beta_{2}\rightarrow1-(1+k_{2}^{L}-h_{2}^{L})(3\rho)/(5\bar{\rho})$,where $k_{2}^{T}$ and $h_{2}^{T}$ are tidal Love numbers and $k_{2}^{L}$ and $h_{2}^{L}$ are load Love numbers. Nominal values assumed for all these parameters are tabulated in Table~\ref{tab:Callisto params}. 

A key feature of the model in \citet{hay2019subsurface} is that it uses bottom drag to account for energy dissipation. This is helpful because as discussed in more detail below and in \cite{hay2017numerically}, the bottom drag coefficient $c_D$ is known, at least approximately, for terrestrial oceans. This is in contrast to some alternative parameterizations of bottom friction. What we find for obliquity tides in the bottom drag scenario is that there is a trade-off between drag and energy dissipation (Fig.~\ref{fig:taucd}). If drag is unimportant, flow velocities will be uninhibited but the effective viscosity will be low, resulting in less energy dissipated and a linear increase in dissipation with viscosity. On the other hand, if drag is important, viscosity will be large, but the flow velocities will be reduced, and the energy dissipation will decrease again \citep[see also][Fig. 3]{chen2014tidal}. Physically, whether bottom drag affects the velocity is determined by the Reynolds number $\Omega R^2/\nu_{\text{obl}}$; the other important dimensionless quantity in Equation~\ref{Edot_subocean} is the Lamb parameter $4 \Omega^2 R^2 / gh$ which denotes the relative speeds of surface gravity waves compared to rotation \citep{chen2014tidal}.

The biggest uncertainties in these analytical expressions are the bottom drag coefficient $c_D$, the ocean thickness $h$, and the factors that encapsulate the effect of the rigid ice shell, $\beta_2$ and $\upsilon_2$. Below, we will explore the sensitivity of Callisto's inclination damping time-scale to the uncertainties in our knowledge of these parameters. 

\subsection{Drag coefficient at the bottom of the ice shell}

\begin{figure}
     \centering
     \subfloat[Bottom drag coefficient\label{fig:taucd}]{
         \includegraphics[width=.32\textwidth]{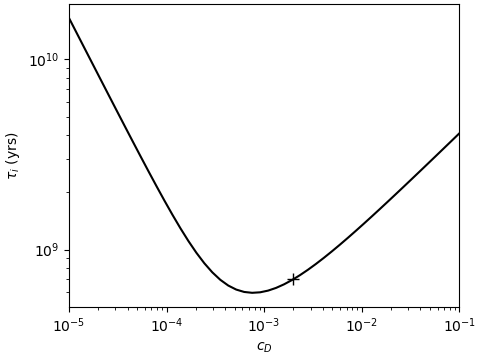}}
     \hfill
     \subfloat[Ocean thickness\label{fig:tauh}]{
         \includegraphics[width=.32\textwidth]{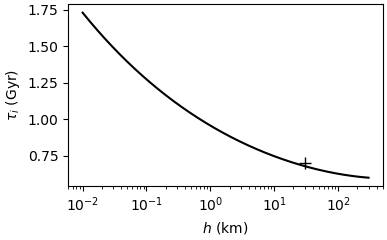}}
          \hfill
     \subfloat[Ice shell thickness\label{fig:taud}]{
         \includegraphics[width=.32\textwidth]{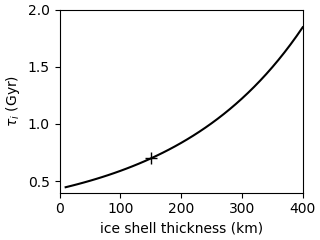}}
        \caption{How Callisto's inclination lifetime varies with physical parameters using Eq.~\ref{Edot_subocean}. Values used in the rest of the paper are marked with a "+" and are $c_D=0.002$, $h=30$~km, and $d=150$~km. For these values $\beta_2$=0.88 and $\upsilon_2$=1.05.}
        \label{fig:tau phys params}
\end{figure}

In \cite{hay2019subsurface}, all of the dissipation in the ocean is modelled as friction at the ocean floor. The bottom drag coefficient estimate, $c_D$, is 0.002, the commonly-assumed value for oceans on Earth that is often deemed applicable to other bodies as well \citep[e.g., ][]{jeffreys1925lxxxiv, sagan1982tide, sohl1995tidal, hay2017numerically}. Fig.~\ref{fig:taucd} plots the inclination damping time-scale as a function of $c_D$, showing that this value would have to be two orders of magnitude larger or smaller than on Earth for the inclination lifetime to become comparable to the age of the solar system.

To investigate the value of $c_D$ further, we use the empirically-derived expression for bottom drag from \cite{turcotte1982application} that depends on the Reynolds number to see whether Callisto's $c_D$ could be two orders of magnitude larger or smaller than on the Earth:
\begin{equation}\label{fb}
c_D = 0.3164\left(\frac{\rho v h}{\eta}\right)^{-1/4},
\end{equation}
\noindent
where $\rho$ is the subsurface ocean density, $v$ is the flow speed, $h$ is the ocean thickness, and $\eta$ is the molecular viscosity \citep{turcotte1982application}. We adopt values consistent with liquid water, $\rho=1000$ kg m$^{-3}$ and $\eta=10^{-3}$ Pa s \citep{sohl1995tidal}. In \cite{chen2014tidal}, the flow speed is a function of effective viscosity, which itself depends on the bottom drag coefficient, so we can simultaneously solve for $v$ and $c_D$ given a specific ocean thickness.

Fig.~\ref{fig:tauh} combines the obliquity tide ocean dissipation equations from Eq.~\ref{Edot_subocean} with the drag coefficient relation from Eq.~\ref{fb} to plot Callisto's inclination damping time-scale as a function of ocean thickness. Flow speed varies with every calculated point and is determined by the expressions in Table~4 of \citet{chen2014tidal}. For a nominal ocean thickness of 30~km, $c_D = 0.0018$, which is almost exactly the standard value used of $c_D = 0.002$. When $h=10$~m, $c_D=0.017$, and when $h=200$~km, $c_D=0.001$. The velocity always stays around a few~cm~s$^{-1}$, and $c_D$ stays within an order of magnitude of the nominal value. Callisto's ocean would have to be less than 10~m thick for drag to be weak enough to lengthen the inclination lifetime to 1.6~Gyr (still too short). Even then, for such a thin ocean, resonances (not captured by \ref{Edot_subocean}) can arise, increasing dissipation and decreasing the inclination lifetime.

\subsection{Effect of a thick ice shell}
The amount of energy dissipated in Callisto's putative subsurface ocean depends on how rigid and thick the overlying ice shell lid is. If the lid is sufficiently weak, then the ocean would be free to flow and dissipate energy as if there were no lid. If the lid deforms not at all, then there would be nowhere for the ocean to flow and no energy dissipation. The ice shell thickness factors into the quantities $\beta_2$, $\upsilon_2$, and $r_t$ in Eq.~\ref{Edot_subocean}, which capture shell pressure, self-gravity, and solid deformation and are related to the pressure and tidal Love numbers.

There are different estimates for Callisto's surface ice thickness, ranging from 100-300~km. \cite{zimmer2000subsurface} conclude that the ocean starts at less than 200-300~km depth. \cite{moore2003tidal} calculate that the ocean should be centred at 166~km. \cite{mckinnon2006convection} takes the overlying ice thickness to be 180~km, and finally, \cite{kuskov2005internal} take it to be 135-150~km. We assume that Callisto's ice thickness is 150~km, satisfying all of these estimates.

To quantify the effect of the ice shell, we compute $\beta_2$ and $\upsilon_2$ as a function of ice shell thickness in the case of a 30~km thick ocean, and we use the obliquity tide ocean dissipation equations from Eq.~\ref{Edot_subocean} to plot the inclination lifetime in Fig.~\ref{fig:taud}. The inclination damping time-scale remains under a few billion years for ice shell thicknesses smaller than the 300~km maximum determined by Callisto flybys \citep{zimmer2000subsurface}. In reality, the rigid shell thickness may be significantly less than the total shell thickness since the base of the shell is likely warm enough to lack rigidity at tidal frequencies. The result is that it is even less likely for the ice shell to restrict ocean dissipation enough to preserve Callisto's inclination over the lifetime of the solar system.

\subsection*{Summary}

Callisto's inclination damping time-scale is much shorter than the age of the solar system for nominal parameter values. Although in principle a sufficiently rigid lid or a very low drag coefficient could mitigate this problem, in neither case do the parameter values required appear to be realistic.

\section{Increasing Callisto's inclination}
\label{section:fuller}

Without an internal tidal solution to Callisto's non-zero inclination, we are left to find explanations external to Callisto. To do so, we need to look at Callisto's dynamical history. Recent astrometry observations suggest that the satellites of the outer planets could be migrating away from their planets faster than previously expected \citep{lainey2017new, lainey2020resonance}. Semi-major axis migration is mostly driven by tidal dissipation in the planet, which is described by the tidal quality factor, $Q$. Dissipation in the satellite con also change the semi-major axis but this effect is generally much smaller. Standard tidal theory assumes that $Q$ is constant for a planet, whereas the new observations suggest a different and time-variable $Q$ for each modal frequency of the planet's internal structure excited by a different satellite. To provide a physical explanation for this phenomenon, \cite{fuller2016resonance} proposed that giant planet tides are satellite dependent and their interiors evolve just as those of stars do. In the rest of this work, we examine the implications of applying this new tidal theory to tides raised on Jupiter by Callisto using a simple proof-of-concept approach. Given the large uncertainties involved, we leave more detailed treatments for future work.

\subsection{Fuller Model}

\begin{table*}
\centering
\caption {Dynamical parameters used in resonance-locking scenario. } 
\begin{tabular}{ l l l l l l l l l }
\hline
 Body & $a$ ($\times 10^6$~m) & $e$ & $i (^\circ)$ & $\theta_0 (^\circ)$ & $\Omega$ (rad s$^{-1}$) & $p'$ & $t_{\alpha}$ (Gyr) & $t_{\text{tide}}$ (Gyr) \\
 \hline
 Jupiter & 778.57$\times 10^3$ & - & - & - & 1.76$\times 10^{-4}$ & - & - & - \\
 Io & 421.8 & - & - & - & 4.11$\times 10^{-5}$ & - & 44 & 20 \\
 Europa & 671.1 & - & - & - & 2.05$\times 10^{-5}$ & - & 101 & 20 \\
 Ganymede & 1070.4 & 0.0094 & 0.177 & -0.05 & 1.02$\times 10^{-5}$ & -6972 & 217 & 20 \\
 Callisto & 1882.7 & 0.0074 & 0.192 & -0.3 & 4.36$\times 10^{-6}$ & -11423 & 63 & 2.7\\  
\end{tabular}
\label{tab:Fuller phys}
\end{table*}

\begin{table}
\centering
\caption {Physical parameters used in resonance-locking scenario} 
\begin{tabular}{ l l l l l l }
\hline
 Body & $M$ ($\times 10^{20}$kg) & $R$ (km) & $J_2$ & $C_{2,2}$ & $c$ \\ 
 \hline
 Jupiter & 1.898$\times 10^7$ & 71398 & 1.474$\times 10^{-2}$ & - & - \\
 Io & 893.2 & - & - & - & - \\
 Europa & 480.0 & - & - & - & - \\
 Ganymede & 1481.9 & 2631.2 & 10/3$C_{2,2}$ & 3.83$\times 10^{-5}$ & 0.311 \\
 Callisto & 1075.9 & 2410.3 & 10/3$C_{2,2}$ & 1.02$\times 10^{-5}$ & 0.353 \\
\end{tabular}
\label{tab:Fuller dyn}
\end{table}

A satellite orbiting a planet will raise a tidal bulge on the planet just as the planet will raise a tidal bulge on the satellite. As the planet rotates faster than the satellite orbits, the planet's bulge will be ahead of the line connecting the planet to the satellite. The satellite will have a net torque on the planet in an effort to slow down the planet's spin to realign the bulge with the direction vector to the satellite. Angular momentum gets transferred from the spin of the planet to the orbit of the satellite, and the lost spin rotational kinetic energy of the planet is dissipated in its interior. 

The model proposed in \cite{fuller2016resonance} connects the phase lag between the planet's tidal bulge and the direction vector to the satellite, which is governed by the difference between the planet's spin rate and the satellite's orbit rate, to the eventual dissipation in the planet. The dissipation could also be thought of as internal friction between the bulge and the rest of the planet as it tries to reposition itself. \cite{fuller2016resonance} presume that there are certain resonant frequencies arising from the planet's structure at which dissipation is greatly increased. If the orbital frequency of the satellite in the planet's spin rotational reference frame is the same as one of these resonant modes, then there will be enhanced dissipation and a greater transfer of angular momentum to the satellite's orbit, forcing it to migrate outwards faster. The final step in \cite{fuller2016resonance} is to suggest that the frequencies of these resonant modes evolve at a time-scale set by the evolution time-scale of the planet; this same time-scale then sets the rate of outwards evolution of the satellite. In this scenario, the satellite's orbital frequency is in resonance with the mode of the structure and gets locked into that resonance hence the name of the phenomenon, resonance locking.

For a mode in Jupiter's interior that evolves over a time-scale $t_\alpha$, to which Callisto's orbital frequency is resonantly-locked, Callisto's orbital frequency or mean-motion, $n$, as a function of time from \cite{nimmo2018thermal} is
\begin{equation}
\label{nimmon}
    n(t) = \Omega_p+\left(n_{\text{now}}-\Omega_p\right)\exp{\left(\frac{t-t_{\text{now}}}{t_\alpha}\right)},
\end{equation}
\noindent
where $\Omega_p$ is the spin frequency of the planet and is assumed constant, $n_{\text{now}}$ is the present-day mean-motion, and $t_{\text{now}}$ is time at present. $t_\alpha$ could be different for each satellite orbiting a planet because each satellite would be resonantly-locked to different interior structures. Astrometry of a satellite's migration rate $da/dt$ can tell us the hypothetical values of $t_\alpha$ by the following approximate relation \citep[][Eq. 12]{fuller2016resonance} in the limit where the planet's spin decays more slowly than its interior evolves

\begin{equation}
\label{ttide}
    \frac{1}{t_{\text{tide}}} \equiv \frac{1}{a}\frac{da}{dt} = \frac{2}{3}\frac{1}{t_\alpha}\left(\frac{\Omega_p}{n}-1\right).
\end{equation}
\noindent
The quantity $t_{\text{tide}}$ represents the time it would take the orbital energy stored in the orbital distance to change by a factor of itself. By definition of $t_{\text{tide}}$ in Eq.~\ref{ttide}, satellites in a mean-motion resonance (MMR), where the orbital periods of two or more satellites are integer multiples of each other, have the same value of $t_{\text{tide}}$. Io, Europa, and Ganymede are in the Laplace resonance, i.e., the 4:2:1 three-body MMR. \cite{fuller2016resonance} and \cite{lainey2009strong} estimate their $t_{\text{tide}}$ at the present-day to be 20~Gyr and Callisto's, which is not part of the resonance, to be $\sim 2$~Gyr. The error bars on the Laplace resonance $t_{\text{tide}}$ are a few Gyr \citep{lainey2009strong, fuller2016resonance}, and there are as yet no astrometry measurements for Callisto. The present-day $t_{\text{tide}}$ values are used to derive the satellites' long-term $t_\alpha$ values using Eq. \ref{ttide}.

\begin{figure}
    {\includegraphics[width=\columnwidth]{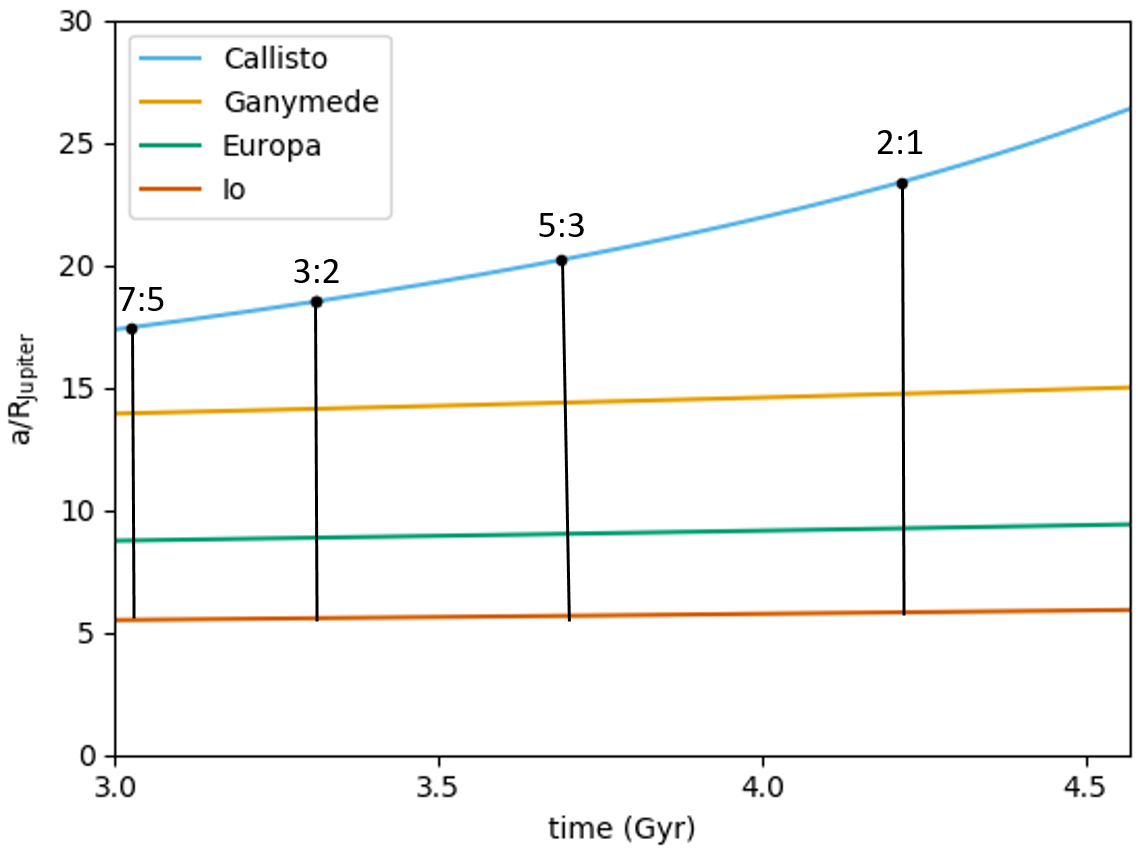}}%
    \caption{Semi-major axis evolution for the Galilean satellites over the past 1.5~Gyr assuming a resonance locking scenario as in \protect\citet{fuller2016resonance}. The $t_\alpha$ values for Io, Europa, Ganymede, and Callisto are 44, 101, 217, and 70~Gyr. The corresponding present-day $t_{\text{tide}}$ values are 20~Gyr for the Laplace resonance and 2.7~Gyr for Callisto. To justify the time span shown, it is assumed that the Laplace resonance began before 3~Gyr. Nominal locations of low-order resonances between Callisto and Ganymede are indicated.}%
    \label{fig:fulleras}%
\end{figure}

Fig.~\ref{fig:fulleras} plots an example of what the resonance locking semi-major axis evolution would look like for the Galilean satellites over the past 1.5 Gyr where at the present-day, t$_{\text{tide}}$ = 2.7~Gyr for Callisto and 20~Gyr for the inner satellites. Callisto's t$_{\text{tide}}$ is an order of magnitude smaller because Callisto has the smallest mean-motion (see Eq. \ref{ttide}).
There is a characteristic concave-up shape associated with semi-major axis evolution in the resonance locking scenario compared to the concave-down shape in a constant $Q$ one. The $k_2/Q$ of Jupiter at Callisto's orbital frequency varies from $0.007$ at 3~Gyr to $0.2$ at present-day, using Eq.~2 from \cite{fuller2016resonance}. [Note that this equation allows us to determine $k_2/Q$ from dynamical parameters and $t_{\alpha}$ without having to specify $k_2$ and $Q$ individually.] The semi-major axis evolution in Fig.~\ref{fig:fulleras} considers only planet tides. For Callisto, the effect of satellite tides on semi-major axis gets within an order of magnitude of that of planet tides when the element values peak (see below), and for Ganymede, satellite tides get within a factor of two at the peak. These peaks are so short-lived, however, that we can neglect their effects on the long-term outwards motion of the satellites.

The Fuller model suggests that over the past 1.5~Gyr, Callisto has migrated more than $30$ per cent of its current orbital distance. The most dynamically relevant consequence of this is that it would have passed through several locations of MMRs, which are marked. Note that the distance ratio between Io, Europa, and Ganymede never changes by virtue of the Laplace resonance. All of them though are on diverging orbits with Callisto because Callisto is pushed away faster. Diverging orbits mean that Callisto could never get caught in an MMR with the other moons since the planet torques pushing Callisto outwards would be stronger than the resonance torques from the other satellites trying to keep Callisto in an MMR \citep[e.g.,][]{dermott1988dynamics}.

Of course there are uncertainties associated with the migration scenario shown in Fig.~\ref{fig:fulleras}. With regards to Io, Europa, and Ganymede, there are uncertainties in the astrometry data used to get $t_{\text{tide}}$ and from there $t_\alpha$, and we do not know when the Laplace resonance was established. For the purposes of providing an example, Fig. \ref{fig:fulleras} shows the Laplace resonance as being intact for the last 1.5~Gyr. With regards to Callisto, the value of $t_{\text{tide}} \approx 2$~Gyr suggested by \cite{fuller2016resonance} has not yet been confirmed by astrometry, though the example of Titan suggests that rapid motion is possible \citep{lainey2020resonance}. For all moons, we do not know if all are in a resonance lock, if some are, or when it started.

\label{section:mmrs}
\subsection{Mean-motion resonances}
If Callisto has passed through MMRs with the inner Galilean moons, then it is important to understand how these transient resonant torques would have affected Callisto's orbit.
To begin, being in an MMR is able to increase the eccentricity, $e$, or inclination, $i$, of a body in either an eccentricity-type or inclination-type resonance. At low order, resonances can be stabilizing, so once entered, a body would have a hard time exiting the resonance without the help of an external force or chaos driven by resonance overlapping \citep{dermott1988dynamics}. Bodies can only enter an MMR if they are on converging orbits otherwise the stronger tidal planetary torque on the outer body causing the orbits to diverge would be enough to disrupt the MMR. Since Callisto is on a diverging orbit with the inner Galilean moons in the Fuller Model, it would have passed through MMRs without getting caught. 

Crossing through resonances would have excited Callisto's eccentricity or inclination \citep{peale1986orbital}. \cite{dermott1988dynamics} give equations for the amount that the orbital elements increase by upon passage through first- and second-order resonances. The details are worked through in \cite{peale1986orbital} and \cite{murray1999solar} to arrive at the third-order resonance as well. The equations are derived by calculating the Hamiltonian of the system before and after entering a resonance. These theoretical expressions are then tested numerically in \citet{dermott1988dynamics} to understand if and when the theory holds up. They find that in systems where all second-order resonances are well-separated, the theory is valid. Two resonances are well-separated if the distance between them is greater than the sum of half their resonance widths. If there is overlap, \citet{dermott1988dynamics} find that satellites can hop between resonances and experience chaos, but they still sometimes obey the resonance crossing theory. The latter is true in particular for the less massive satellite and for eccentricity resonances more than inclination ones. The authors argue that the theory applies when the rate of semi-major axis oscillation in a resonance (i.e., librating within the resonance width in one period) is faster than the tidal semi-major axis migration rate. Callisto satisfies this latter criterion, even though some of its resonances are not well-separated; we discuss this issue further in Section~3.3 below.

The equations for the final eccentricity or inclination of the inner body after passing through a $q$-order resonance, $x_q$, are below and assume that the element value before the resonance crossing is zero:
\begin{equation}
\label{xcrit}
    \begin{split}
        \displaystyle x_1 &= \displaystyle\left[\frac
{2\sqrt{6}f(\alpha)\left(m'/M\right)\alpha}
{p^2+(p+1)^2\left(m/m'\right)\alpha^2}
\right]^{1/3}\\
\displaystyle x_2  &= \displaystyle \left[\frac
{\left(32/3\right)f(\alpha)\left(m'/M\right)\alpha}
{p^2+(p+2)^2\left(m/m'\right)\alpha^2}
\right]^{1/2} \\
\displaystyle x_3  &= \displaystyle\frac
{9\sqrt{2}f(\alpha)\left(m'/M\right)\alpha}
{p^2+(p+3)^2\left(m/m'\right)\alpha^2}.\\
    \end{split}
\end{equation}
\noindent
Each equation is for a $p:p+q$ resonance. The unprimed $m$'s and $a$'s are the mass and semi-major axis for the inner body, and the primed ones are for the outer body. $M$ is the central body's mass. $\alpha = a$/$a'$ is the ratio of the inner body's semi-major axis to the outer body's. $f(\alpha)$ is a function of Laplace coefficients commonly used in Hamiltonian mechanics \citep{murray1999solar} and is different depending on the order of the resonance and the type (i.e., eccentricity vs. inclination and inner body vs. outer body). For $q>1$, we make the same assumption that \cite{dermott1988dynamics} do that the final element values hold true for mixed resonances (e.g., $ii$') as well as pure ones (e.g., $i^2$, $i'^2$) so long as the correct $f(\alpha)$ is used.

The final element values for the outer body after passing through a $p:p+q$ resonance have the same form except every factor of $\alpha$ outside of the Laplace coefficients becomes $\alpha^3$, every factor of $\alpha^2$ becomes $\alpha^4$, and each $m$ and $m'$ is swapped for the other.

We note that in \cite{dermott1988dynamics} it is stated that $e_q = 2 i_q$. However, \cite{murray1999solar} claim that $e_q = i_q$ because the derivations of both values are the same whether one starts with eccentricity or inclination angular momentum. Here we follow the convention of \cite{murray1999solar} and use the same equations to calculate both $e_q$ and $i_q$.

Applying these equations to Callisto's inclination, we ask: do there exist resonances with the inner Jovian satellites, Io, Europa, and Ganymede that could have excited Callisto's inclination? Note that inclination-type resonances are only possible in even-ordered resonances or mixed resonances that have pairs of inclination terms, so we will only consider $p:p+2$ resonances, $i^2, ii', i'^2$ \citep[][Section 8.4]{murray1999solar}. A conceptual reason for this is that there is no standard reference frame that appears in the physics, so only the mutual inclination matters. There are two types of inclination-type resonances that affect the inclination of Callisto, $i-$Ganymede-$i-$Callisto and $i^2-$ Callisto, denoted $ii', i'^2$. 

\begin{figure}
     \centering
     \subfloat[Inclination boosts for second-order resonances since inclination-type resonances only occur for even-orders.\label{fig:iboosts}]{
         \includegraphics[width=.48\textwidth]{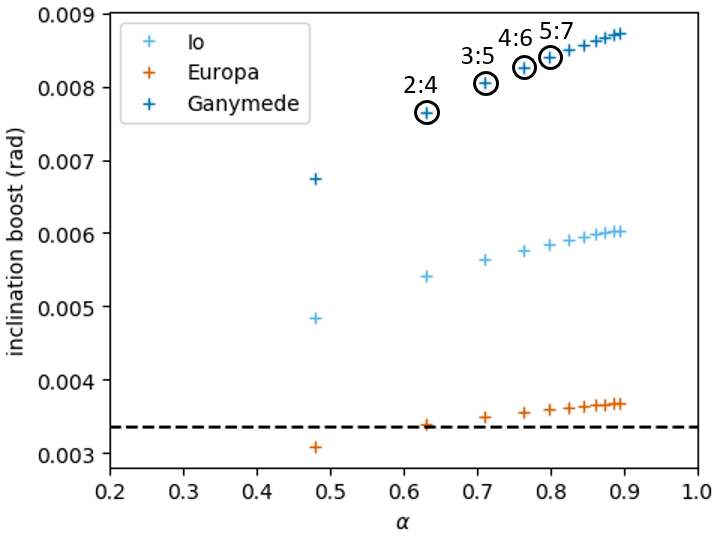}}
     \hfill
     \subfloat[Eccentricity boosts for first-order (square), second-order (diamond), and third order (circle) resonances.\label{fig:eboosts}]{
         \includegraphics[width=.48\textwidth]{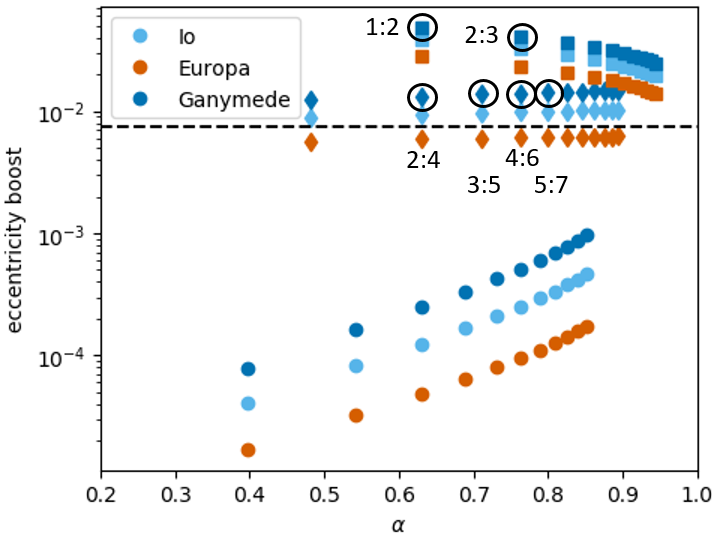}}
    \caption{Callisto's element boosts after passing through $p:p+q$, $1\le p \le 11$ resonances with Io (light blue), Europa (dark orange), and Ganymede (dark blue). In exact resonance, $\alpha = \left[p/(p+q)\right]^{2/3}$, so $\alpha$ approaches 1 with increasing $p$. The second- and third-order resonances here are mixed $x^{q-1}$-Moon-$x-$Callisto resonances because they produce higher element boosts. The resonances that Callisto passes through in Fig.~\ref{fig:fulleras} with Ganymede are circled and labelled. The dashed lines are Callisto's current $i$ and $e$.}
        \label{fig:element boosts}
\end{figure}

For each of Io, Europa, and Ganymede, we calculate $x_2$ from Eq.~\ref{xcrit} for $1\le p \le 11$, plotted in Fig.~\ref{fig:iboosts}. All resonances but one increase Callisto's inclination to at or above its current value by a factor of 2-3. Theoretically, therefore, a resonance crossing could be responsible for Callisto's inclination. 

Callisto's eccentricity has not been mentioned thus far because solid-body eccentricity tidal damping depends on its value of $k_2/Q$, which is unknown. However, if Callisto had passed through a second-order inclination-type resonance with Ganymede, then it would have soon thereafter passed through eccentricity-type resonances as well. Fig.~\ref{fig:eboosts} shows first, second, and third-order eccentricity boosts, of which the first and second-order ones could account for Callisto's present-day eccentricity. The next section will provide a more detailed investigation of whether the present-day eccentricity and inclination can be explained by passage through a resonance.

\subsection{Inclination evolution}

Assuming the semi-major axis migration model for the Galilean satellites from \cite{fuller2016resonance} (Section~3.1), we can track Callisto's inclination as it crosses MMRs with Ganymede and subsequently decays due to obliquity tides.

We will assume that inclination and obliquity are connected through the Cassini state relation \citep{ward1975past}, generally appropriate for dissipative systems. For a synchronously rotating satellite, this says that the spin pole, orbit normal, and Laplace plane normal vectors all lie in the same plane. The Cassini state relation is
\begin{equation}
\label{cassini}
    \frac{3}{2}\left[\left(J_2+C_{2,2}\right)\cos{\theta_0}+C_{2,2}\right]p'\sin{\theta_0} = c'\sin{\left(i-\theta_0\right)},
\end{equation}
where $J_2$ and $C_{2,2}$ are the degree-2 gravity coefficients, $\theta_0$ is obliquity, $c'$ is an effective normalised moment of inertia, $i$ is inclination, $p' = \displaystyle \Omega /\dot{\Omega}_{\text{orb}}$, $\Omega$ is spin frequency, and $\dot{\Omega}_{\text{orb}}$ is the precession rate of the longitude of the ascending node \citep[e.g.,][]{chen2014tidal}. 

Titan is the only icy satellite whose obliquity has been directly measured \citep{stiles2008determining}, and for the Cassini State relation to hold in Titan's case, $c' = 1.9c$, where $c$ is the usual normalised moment of inertia, as derived from gravity moments and the hydrostatic assumption \citep{bills2011rotational}. The result, a normalised moment of inertia above 0.4, suggests that Titan's ice shell and interior are decoupled due to a subsurface ocean \citep{bills2011rotational,baland2011titan}. Since Callisto, like Titan, is believed to have a subsurface ocean, it is reasonable to assume that $c' > c$. For specificity, below we will assume $c'=1.6c$ (implying a present-day obliquity of $-0.3^\circ$ instead of $-0.12^\circ$ for $c'=c$), though of course the actual relationship is currently unknown. We note that \citet{baland2012obliquity} predict an obliquity as large as $-0.25^\circ$ for a decoupled shell, while for a solid Callisto, \citet{bills2005free} and \citet{noyelles2009expression} obtain values as large as $-1.7$ and $-0.27^\circ$ respectively. As the numerical factor in front of $c$ increases, obliquity will be larger and generate stronger obliquity tides for a given inclination. Stronger obliquity tides would then damp out inclination faster. 

Using the measured value of $C_{2,2}$ from \cite{anderson2001shape}, we derive $c = 0.353$ from the Darwin-Radau relation, which assumes hydrostatic equilibrium:

\begin{equation}
\label{C22}
    C_{2,2} = \frac{1}{4}\left(\frac{\Omega^2}{\frac{4}{3}\pi \rho_{\text{sat}}G}\right)\left(\frac{5}{1+\left(\frac{5}{2}-\frac{15}{4}c\right)^2}-1\right),
\end{equation}
where $\Omega$ is spin frequency, $\rho_{\text{sat}}$ is the satellite's bulk density, and $G$ is the gravitational constant.

Callisto is in a synchronous rotation state, so the spin frequency is the same as the orbital frequency and decreases as it migrates outwards. Making the hydrostatic assumption means that as Callisto's spin frequency changes, the amount of spherical flattening will change as well. $J_2$ and $C_{2,2}$ must therefore be calculated at every time step in Callisto's outwards migration. Note that $J_2$ and $C_{2,2}$ have not been independently measured for Callisto, so we do not know that the hydrostatic assumption is correct, but they can be determined assuming that Callisto is hydrostatic via $J_2 = 10 C_{2,2}/3$.
 
The $\Omega_{\text{orb}}$ circulates due to the additional gravitational force arising from Jupiter's oblateness, the inner Galilean moons, and the Sun, which is treated as an external perturber. The way we calculated Callisto's $\Omega_{\text{orb}}$ precession is as follows. The terms are either from internal or external perturbers as per the Lagrange equations for planetary motion \citep[e.g.,][]{champenois1999chaos, noyelles2009expression}.
\begin{equation}
\label{nodalprecession}
\begin{split}
    \frac{d\Omega_{\text{orb}}}{dt} = -\frac{3}{2}n_4J_{2p}\left(\frac{R_p}{a_4}\right)^2-&\sum_{i=1}^{3}\frac{1}{4}n_4\alpha_{i,4}\frac{m_i}{M}b_{3/2}^{(1)}(\alpha_{i,4})\\
    -&\frac{1}{4}n_4\alpha_{4,s}^2\frac{m_s}{M}b_{3/2}^{(1)}(\alpha_{4,s})
\end{split}
\end{equation}
The subscripts 1-3 refer to Io, Europa, and Ganymede respectively, $p$ refers to the planet, namely Jupiter, and $s$ refers to the Sun. $J_{2p}$, $R_p$, and $M$ are Jupiter's degree-2 gravity coefficient, radius, and mass. $n_4$ and $a_4$ are Callisto's mean-motion and semi-major axis. $m_i$ is the $i^{\text{th}}$ object's mass. $\alpha_{i,j} = a_i/a_j$ is the ratio of the i$^{\text{th}}$ object's semi-major axis to the j$^{\text{th}}$ object's. $b^{(j)}_{s}(\alpha)$ are Laplace coefficients as a function of $\alpha$ \citep{murray1999solar}. With Jupiter at the centre of the system, the semi-major axis of the Sun is just the heliocentric semi-major axis of Jupiter. In the case of Callisto, precession due to Jupiter's oblateness is smaller than precession due to Ganymede, unlike all other Galilean moons whose precessions are dominated by Jupiter's oblateness.

Regarding the effects of tidal dissipation, we take into account the fact that dissipation in the primary decreases a satellite's inclination as well as dissipation in the satellite \citep{chyba1989tidal}. The Mignard equations for how the Moon's orbital elements evolve due to dissipation in the primary and secondary consider only solid-body satellite tides \citep{mignard1981lunar}. We use the adaptation in \cite{chen2016tidal} to include the effect of ocean satellite tides in addition to solid-body tides:
\begin{equation}
\label{mignard1}
    \begin{split}
        \displaystyle \left(\frac{di}{dt}\right)_\text{p} &= \displaystyle -\frac{3}{4}\frac{k_{2,p}}{Q_p}\left(\frac{R_p}{a}\right)^5\frac{m}{M}n\sin{i}\frac{\Omega_p}{\Omega_p-n}\sqrt{1+\frac{m}{M}}\cos{\theta_{0,p}}\\
        \displaystyle \left(\frac{di}{dt}\right)_\text{s}  &= \displaystyle -\frac{a\dot{E}_{\text{obl}}}{GMm\tan i}\sqrt{1+\frac{m}{M}}\\
        \displaystyle \frac{di}{dt} &= \left(\frac{di}{dt}\right)_\text{s} + \left(\frac{di}{dt}\right)_\text{p},
    \end{split}
\end{equation}
\noindent
where the subscript $p$ denotes quantities for the planet and the subscript $s$ for the satellite. $\dot{E}_{\text{obl}}$ includes both ocean (Eq.~\ref{Edot_subocean}) and solid-body (Eq.~\ref{diss solid}) dissipation in the satellite.

Note that in the limit where $n \ll \Omega_p$, $m\ll M$, and $\cos{\theta_{0,p}=1}$, $(di/dt)_p$ simplifies to the expression in \cite{chyba1989tidal} and $(di/dt)_s$ simplifies to the expression in \cite{chen2016tidal}.

Eq. 2 of \cite{fuller2016resonance} calculates $k_{2,p}/Q_{p}$ at Callisto's orbital frequency and is a measure of how deformable Jupiter is :
\begin{equation}
    \frac{k_{2,p}}{Q_{p}} = \frac{1}{3n}\frac{M}{m}\left(\frac{a}{R_p}\right)^5\frac{1}{t_{\text{tide}}}.
\end{equation}

The four most recent resonances with Ganymede going backwards in time are 2:1, 5:3, 3:2, and 7:5 (the same resonances with Europa are 4:1, 10:3, 3:1, and 14:5) as shown in Fig.~\ref{fig:fulleras}.
Second-order inclination resonances can either excite just the inner body's inclination ($i^2$), just the outer body's inclination ($i'^2$), or both ($ii'$). We consider the sum of the second-order inclination resonances that excite Callisto's inclination ($i'^2$ and $ii'$). Even though the second-order resonances are not well-separated, at least for the most recent 2:1 crossing, \citet{dermott1988dynamics} note that there are times when the resonance crossing theory still holds even for poorly-separated resonances (see the discussion in section \ref{section:mmrs}). They speculate that the theory could still hold for the less massive satellite in a pair and if the semi-major axes of the satellite are expanding adiabatically (i.e., the tidal semi-major axis migration rate is slower than the resonant semi-major axis libration rate). Callisto and Ganymede are roughly the same mass, and the tidal expansion rate is much slower than the resonant libration rate. These are not guarantees, of course; understanding in detail which second-order resonances would excite Callisto's inclination would only be possible with N-body simulations that take into account all of the necessary resonant physics. That is outside the scope of this work.

We assume that Callisto acquires the inclination boost over a finite amount of time and not instantaneously. We approximate that the inclination increases exponentially as Callisto approaches the exact value of the resonance (this assumption has not been tested numerically, but correctly reproduces the total orbital element boost - see below). The resonance crossing time-scale, $\tau_{\text{res}}$, is the time it takes Callisto's and Ganymede's motions to cross the full resonance width:
\begin{equation}
\label{resonance width}
    \tau_{\text{res}} = \frac{\Delta a}{\dot{a}} = \frac{2a}{\dot{a}}\left[\frac{16}{3}\left(\alpha\frac{m'}{M}+\frac{m}{M}\right)f(\alpha)z^q\right]^{1/2},
\end{equation}
where $\Delta a$ is the full resonance width, $\dot{a}$ is the migration rate from Eq.~\ref{ttide}, and $z^q$ is either $e^q$ or $\sin(i/2)^q$ \citep{dermott1988dynamics}. The rate of eccentricity and inclination increase as Callisto approaches the resonance is taken to be
\begin{equation}
\label{resonance tau}
    \frac{dx}{dt} = \frac{x_{q}}{2}\frac{1}{\tau_{\text{res}}}\exp(-|t-t_{\text{res}}|/\tau_{\text{res}}),
\end{equation}
where $t_{\text{res}}$ is the time where the exact $p:p+q$ resonance occurs, and $x_{q}$ is given by (\ref{xcrit}). When integrated over time, this expression delivers the correct total increase in orbital element ($x_{q}$).

\begin{figure}
    \centering
{\includegraphics[width=\columnwidth]{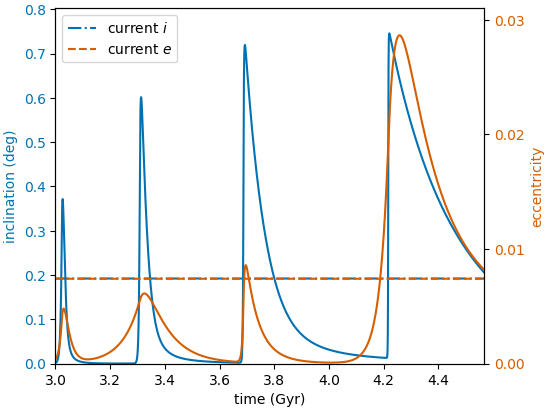}}%
        \caption{Callisto's inclination (left, blue) and eccentricity (right, orange) evolution for $t_\alpha = 70$ Gyr, $d=150$ km, and solid-body $k_2/Q = 0.045$. The inclination boosts are sums of the $i-$Ganymede-$i-$Callisto and $i^2-$Callisto resonances, and the eccentricity boosts are the $e-$Callisto ($e^2-$Callisto for second-order) resonances. This model is able to reproduce Callisto's present-day orbital elements.}%
    \label{fig:e-i-evol}%
\end{figure}

To generate Fig.~\ref{fig:e-i-evol}, we let Callisto's inclination increase as it approaches the resonance (Eq.~\ref{resonance tau}) and decay due to tides raised on the planet and satellite obliquity tides in both a subsurface ocean and in the solid body (Eq.~\ref{mignard1}). All of the physical and dynamical parameters used can be found in Tables~\ref{tab:Fuller phys} and ~\ref{tab:Fuller dyn}. The inclination evolution has three free parameters since the drag coefficient is taken to be 0.002: $t_\alpha$, the resonance locking migration time-scale, $d$, the thickness of the ice shell overlying Callisto's ocean (which controls the dissipation rate), and the solid-body $k_2/Q$, which controls the solid-body obliquity tide dissipation rate. $k_2/Q$ can be solved for, however, using Callisto's eccentricity evolution, which we analyse in the next section. This means for every $t_\alpha$, there is one $k_2/Q$ which allows us to recover Callisto's eccentricity, and that is the $k_2/Q$ that we use for Callisto's solid-body obliquity tides as well. In the example shown in Fig.~\ref{fig:e-i-evol}, where $t_\alpha = 70$~Gyr, $d = 150$~km, and $k_2/Q = 0.045$, the inclination evolution that includes the sum of both resonance excitations matches Callisto's present-day inclination. The predicted final obliquity from the Cassini state relation is $-0.3^\circ$.

\begin{figure}
     \centering
     \subfloat[Inclination\label{fig:i-error}]{
         \includegraphics[width=.48\textwidth]{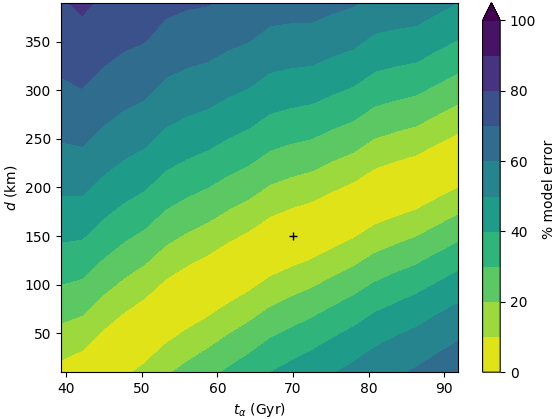}}
     \hfill
     \subfloat[Eccentricity\label{fig:e-error}]{
         \includegraphics[width=.48\textwidth]{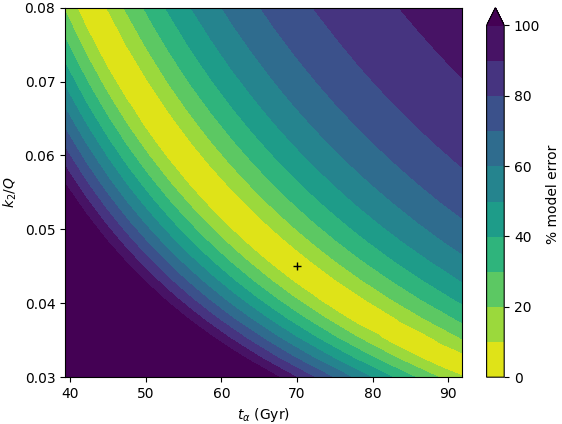}}
    \caption{The error between the inclination (top) and eccentricity (bottom) evolution models and Callisto's present-day values assuming different $t_\alpha$, $d$, and $k_2/Q$. $k_2/Q$ for the solid-body obliquity tide component in the inclination contour plot is fixed for every $t_\alpha$ and is the value used to minimize the eccentricity error in the right plot for a given $t_\alpha$ (i.e., the yellow band). The "+" marks the parameters used in Fig. \ref{fig:e-i-evol}.}
        \label{fig:e-i-error}
\end{figure}

We explored other $t_\alpha$ and $d$ values to determine parameter pairs that could match Callisto's present-day inclination. For different pairs of values, we carried out the full inclination evolution calculation and determined the relative error between the model's final inclination and the actual present-day value. Fig.~\ref{fig:i-error} plots the error contours showing the trade-off between $d$ and $t_{\alpha}$. There is a different solid-body $k_2/Q$ for every $t_\alpha$, and it is constrained by fitting the eccentricity evolution in the next section. Longer migration time-scales require a thicker lid to reproduce Callisto's inclination whereas shorter migration time-scales require a thinner or even absent lid. This is because longer migration time-scales mean that the most recent resonance crossing happened earlier in time, and so thicker lids are needed to hamper ocean dissipation and preserve the inclination for longer.

The accuracy of our results depends on the choice of Callisto's current orbital elements. We have used JPL HORIZONS, which provides Callisto's mean inclination with respect to the local Laplace plane. However, other ways of determining Callisto's exact inclination yield different results, e.g., a frequency analysis approach yields $0.257^\circ$ \citep{noyelles2009expression}. Using as a baseline $0.257^\circ$ instead of $0.192^\circ$ would change our results by about 30 per cent. Given the order-of-magnitude uncertainties in other parameters of interest we are not too concerned by this particular source of uncertainty, though it would obviously be desirable to resolve it in future work.

We also save for future work the task of keeping track of how the Laplace plane of Callisto changes as it migrates away from Jupiter. This would affect the reference plane from which the inclination is measured and thereby change the true value of both the inclination and the obliquity.

\subsection{Eccentricity evolution}
If Callisto is passing through values of semi-major axis that correspond to inclination-type resonances, then if the eccentricity-type resonances are well-separated, we would expect eccentricity boosts as well. We track Callisto's eccentricity evolution to see if Callisto's full orbital history can be made self-consistent. The Mignard equations for eccentricity evolution due to dissipation in the primary and satellite are

\begin{equation*}
    \begin{split}
        &\displaystyle \left(\frac{de}{dt}\right)_\text{p} = \\
        &\displaystyle
        \frac{3}{2}\frac{k_{2,p}}{Q_p}\left(\frac{R_p}{a}\right)^5\frac{m}{M}\left(1+\frac{m}{M}\right)\frac{n^2}{\Omega_p-n}\left(\frac{\Omega_p}{n}\cos{\theta_{0,p}}\cos{i}\frac{f_4}{\beta^{10}}-\frac{f_3}{\beta^{13}}\right)\\
    \end{split}
\end{equation*}

\begin{equation}
\label{mignard2}
    \begin{split}
        &\displaystyle \left(\frac{de}{dt}\right)_\text{s} = \displaystyle
        3\frac{k_{2,s}}{Q_s}\left(\frac{R_s}{a}\right)^5\left(1+\frac{M}{m}\right)n\left(\cos{\theta_{0,s}}\frac{f_4}{\beta^{10}}-\frac{f_3}{\beta^{13}}\right)\\
        &\displaystyle \beta = \displaystyle \sqrt{1-e^2}, \quad \\
        &\displaystyle f_3 = \displaystyle 9e + \frac{135}{4} e^3 + \frac{135}{8} e^5 + \frac{45}{64} e^7, \quad
        \displaystyle f_4 = \displaystyle \frac{11}{2} e + \frac{33}{4} e^3 + \frac{11}{16} e^5\\
        &\displaystyle \frac{de}{dt} = \displaystyle \left(\frac{de}{dt}\right)_\text{s} + \left(\frac{de}{dt}\right)_\text{p},
    \end{split}
\end{equation}
\noindent
where, as above, the subscript $p$ is for planet quantities and $s$ is for satellite quantities. Note that for eccentricity tides, dissipation in the primary increases the eccentricity, which is opposite to its effect on inclination.

In the limit where $n\ll \Omega_p$, $m \ll M$, $\cos{\theta_{0,p}=1}$, and $e\ll 1$, $(de/dt)_p$ simplifies to the expression in \cite{peale1980tidal} adapted from \cite{goldreich1963eccentricity} with a difference in coefficents of about 15 per cent. $(de/dt)_s$ simplifies to the standard solid-body eccentricity decay rate \citep[e.g.,][]{peale1980tidal}. We ignore ocean eccentricity tides because they produce negligible dissipation. To convert the \citet{peale1980tidal} expressions, which assume a constant tidal phase lag, to the Mignard equivalent, which assumes a constant tidal time lag, requires setting $\Delta t_s = 1/(n Q)$ where $Q$ is the tidal quality factor. We also took $\Delta t_p = 1/(2\Omega_p Q)$. Expressions that assume a constant time lag can be made equivalent to expressions that assume a constant phase lag only at one particular frequency. In reality, both $\Delta t$ and $Q$ are expected to be frequency-dependent, but there is little agreement on the nature of this frequency-dependence.

Solid-body eccentricity tides require a $k_2/Q$ for Callisto that is unknown because its internal structure is not well-constrained \citep{moore2003tidal}. That means that for Callisto's eccentricity evolution, the two free parameters are $k_2/Q$ and $t_\alpha$. Since the inclination evolution depends on $t_{\alpha}$, $d$, and $k_2/Q$, we have three unknowns and two constraints. Thus, the coupled inclination-eccentricity problem allows us to determine both $d$ and $k_2/Q$ for a given $t_{\alpha}$. Because of its uncertainty, we here treat the solid-body $k_2/Q$ as a constant (see below).

Fig.~\ref{fig:e-i-evol} shows Callisto's eccentricity evolution as the eccentricity increases with resonance crossings and decays with solid body eccentricity tides. For $t_\alpha = 70$~Gyr, $k_2/Q = 0.045$ to make Callisto's eccentricity evolution match the present-day eccentricity. Eccentricity resonances are first-order where possible, and the rest, including all inclination resonances, are second-order. First-order resonances have a wider resonance width (Eq.~\ref{resonance width}), which explains why for the first and third resonances, eccentricity has a broader peak than inclination. Just as for the inclination evolution, we varied $t_\alpha$ and $k_2/Q$ to find the error between Callisto's present-day eccentricity and the value obtained from our models (Fig.~\ref{fig:e-i-error}). There is an inverse relationship between $t_\alpha$ and $k_2/Q$, so longer migration time-scales require a less dissipative solid-body to reproduce Callisto's eccentricity whereas shorter migration time-scales require the solid-body to be more dissipative.

\subsection{Summary}
It is possible that Callisto's current inclination and eccentricity are remnants of recent resonance crossings between Callisto and Ganymede while both migrated away from Jupiter in a frequency-dependent $Q$ model. We have found combinations of $t_\alpha$, $d$, and $k_2/Q$ that reproduce Callisto's orbital elements within 10 per cent; our nominal values are $t_\alpha = 70$~Gyr, $d = 150$ km, and $k_2/Q = 0.045$ based on likely physical models of Callisto's interior.

It should be noted that tidal dissipation is more efficient the closer Callisto is to the inner satellites and Jupiter, so the only resonance needed to converge on Callisto's current inclination and eccentricity is the most recent one, a 2:1 resonance with Ganymede. Passing through a resonance with Ganymede means passing through a resonance with Europa and Io as well, but even a first-order resonance with Ganymede becomes a third-order resonance with Europa and a fifth-order resonance with Io. Fig.~\ref{fig:eboosts} shows that third-order eccentricity boosts are an order of magnitude below second-order eccentricity boosts, so it can be extrapolated that fourth-order inclination boosts must be several orders of magnitude lower than second-order inclination boosts and therefore can be neglected. As such, we need only to consider Ganymede's perturbation in our model. Since our uncertainty of how long Callisto or the other Galilean moons have been resonantly-locked to Jupiter for and what these migration time-scales are exactly, needing only one resonance in the last $\sim 400$ Myr is an advantage as it means Callisto's uncertain earlier history is not vital to this scenario.

\section{Discussion}
\label{section:discussion}
Even though our focus is on recovering Callisto's orbital elements, different aspects of the system are sure to be affected by Callisto and Ganymede crossing resonances with each other. These aspects need to be consistent with or provide evidence for the story in order for it to hold up. We first analyse the enhanced tidal heating that Callisto experiences as its orbital elements are excited since its surface shows no evidence of reheating. Secondly, we check that Ganymede's orbital elements are consistent with resonance crossings with Callisto. Thirdly, we discuss potential future work and make some testable predictions.

\subsection{Heat flux}
Callisto's surface is old, cratered, and shows no evidence of resurfacing \citep{greeley2000galileo}. Any orbital evolution we propose should satisfy these surface constraints. Too much surface heat flux from tidal dissipation would have potentially relaxed craters, created fractured features, or caused resurfacing. 

\begin{figure}
    \centering
{\includegraphics[width=\columnwidth]{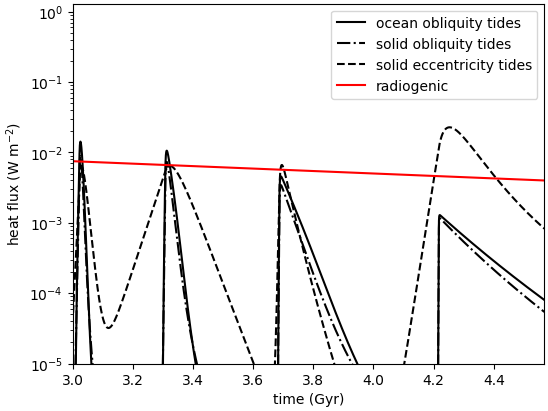}}%
        \caption{Callisto heat flux due to ocean obliquity tides, solid body obliquity tides, and solid body eccentricity tides. Ocean eccentricity tides are negligible. The periodic additions of heat could have prevented Callisto's ocean from freezing out completely in its history. The steady-state radiogenic heat flux estimates are from \protect\citet{mckinnon2006convection}}%
    \label{fig:flux callisto}%
\end{figure}

Fig.~\ref{fig:flux callisto} shows the heat flux for ocean obliquity tides, solid body obliquity tides, and solid body eccentricity tides as Callisto migrates with $t_\alpha = 70$~Gyr. Whether obliquity tides were stronger than eccentricity tides depended on the order of the resonance and how fast Callisto and Ganymede were crossing the resonance. First-order resonances have a wider resonance width, lowering the peak eccentricity heat flux. In the frequency-dependent $Q$ model, resonantly-locked satellites migrate faster at greater semi-major axis, so the resonance crossing time-scale decreases the farther out the resonance is.

The largest heat flux peak is due to eccentricity tides and reaches 30~mW~m$^{-2}$. It is short-lived, dropping down to radiogenic heat levels within $\rm \approx$200~Myr. 
An obvious question is whether a heat pulse of this kind is inconsistent with Callisto's observed lack of geological activity. Perhaps the most important consideration is that the time-scale for a conductive shell to respond to a change in bottom boundary conditions is $d^2/\kappa \approx$700~Myr for $d$=150~km. For a sluggishly-convecting shell as is conceivable with a $\rm 30~mW~m^{-2}$ heat flux \citep{mckinnon2006convection}, this time-scale would be smaller by roughly an order of magnitude. Since the response time-scale is likely comparable to or greater than the heat pulse duration, a heating event at the base of the shell would only be weakly expressed near the surface.

Callisto's large craters in general appear unrelaxed \citep{schenk2002thickness} and it does not possess small ($<$30~km) relaxed craters, unlike Ganymede \citep{singer2018relaxed}. Unfortunately, placing an upper limit on heat flux based on unrelaxed craters is difficult given uncertainties in the relevant ice rheology and basal temperature \citep{bland2017viscous}. 

Given the large uncertainties, a heat pulse peaking at $\rm 30~mW~m^{-2}$ is not obviously inconsistent with the available constraints.  Conversely, convection of Callisto's ice shell would tend to make long-term survival of an ocean more challenging  \citep{reynolds1979internal,mckinnon2006convection} and the addition of intermittent tidal heating might help to explain its present existence. Further investigations of these issues would be of potential interest.

\subsection{Ganymede}
A further reality check is to see whether Ganymede's inclination and eccentricity evolution fit with its having passed through inclination and eccentricity-type resonances with Callisto. This is because if Callisto passed through resonances with Ganymede, then Ganymede would have passed through these resonances as well.

\begin{figure}
    \centering
{\includegraphics[width=\columnwidth]{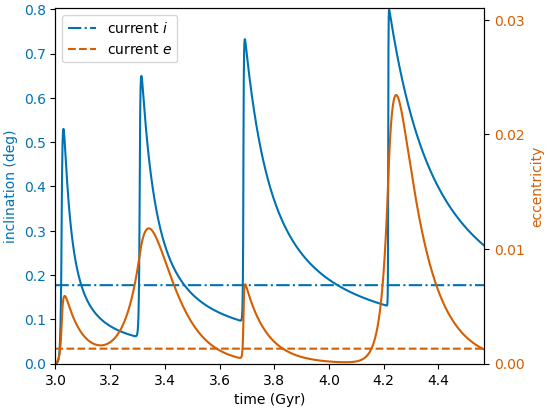}}%
        \caption{Ganymede's inclination (left, blue) and eccentricity (right, orange) evolution for $t_\alpha = 70$~Gyr for Callisto. $t_\alpha = 217$~Gyr, $d=150$~km, $\beta_2=0.85$, $\upsilon_2=1.04$, and $k_2/Q=0.0025$ for Ganymede. The inclination boosts are the sum of the $i^2-$Ganymede and $i-$Ganymede-$i-$Callisto resonances, and the eccentricity boosts are the $e-$Ganymede ($e^2-$ Ganymede for second-order) resonances. Unlike for Callisto, Ganymede's inclination does not always zero out in between resonance crossings. Ganymede's inclination and eccentricity can also be accounted for by resonance crossings with Callisto.}%
    \label{fig:gany e-i}%
\end{figure}

\begin{figure}
    \centering
{\includegraphics[width=\columnwidth]{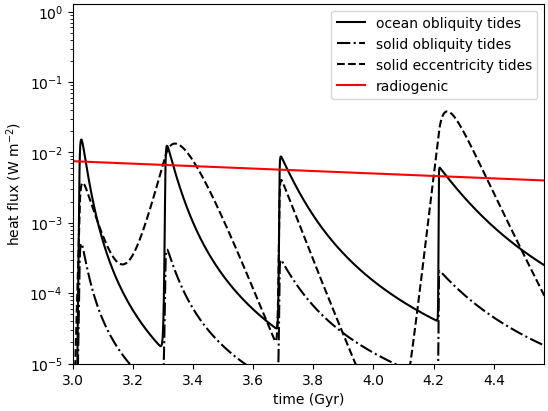}}%
        \caption{Ganymede's heat flux evolution for ocean obliquity tides and solid-body eccentricity tides as it passes through resonances with Callisto. Radiogenic heat flux estimates are from \protect\citet{bland2009orbital}.}%
    \label{fig:ganyheat}%
\end{figure}

What we find is that Ganymede's inclination and eccentricity evolution lines up within a factor of two with the Fuller model when for Callisto $t_\alpha = 70$~Gyr and for Ganymede $t_\alpha = 217$~Gyr, $d=150$~km,$\beta_2=0.85$, $\upsilon_2=1.04$, and $k_2/Q = 0.0025$ (Fig.~\ref{fig:gany e-i}). In contrast with Callisto, Ganymede's ocean obliquity tide dissipation is almost independent of ice shell thickness, because of how much faster it spins. Ganymede's periodic heat flux from tidal dissipation peaks at 45 mW m$^{-2}$ in Fig.~\ref{fig:ganyheat}, which is higher than Callisto's peak heat flux despite having a $k_2/Q$ an order of magnitude lower, because of its smaller semi-major axis. The inference that Ganymede's solid-body $k_2/Q$ is so much lower than that of Callisto is somewhat surprising. One possible, although speculative, explanation is that a partially-differentiated Callisto could result in a large volume of mixed ice and rock having a viscosity dominated by the weaker phase. Because ice-like viscosities are much more readily subject to tidal heating than rock-like viscosities, the overall result would be enhanced dissipation in Callisto.

A complication is that Ganymede experienced a partial resurfacing event mid-way through its history \cite{bland2009orbital}. The most likely explanation for this event is that it spent time in a Laplace-like resonance \citep{showman1997tidal} prior to entering the current Laplace resonance; this might also explain Ganymede's current eccentricity, but would require a $k_2/Q$ at least an order of magnitude lower than we propose. In our model Ganymede could have experienced a heat-pulse as recently as 0.3~Ga (Fig. \ref{fig:ganyheat}), but it is not clear that this heat pulse is sufficient to explain the inferred peak flux in excess of $\rm 100~mW~m^{-2}$ \citep{bland2009orbital}.

Recovering Ganymede's orbital elements is more complicated than for Callisto because of the Laplace resonance with Io and Europa. At the present-day, its longitude of pericentre and longitude of the ascending node are not included in the Laplace resonance angles \citep[e.g.,][]{showman1997tidal}, but if its inclination and eccentricity had gotten large enough in the past, the resonance angles may have been different. In such a case, Ganymede's elements may have been forced to non-zero values for a period of time. Because of the Laplace resonance, the orbital dynamics of Ganymede are a more difficult problem than Callisto, and a full treatment is outside the scope of this work. At present, all we can say is that resonance crossings with Callisto are able to explain Ganymede's elements to within a factor of two.

\subsection{Predictions and Future Work}
In this work we have made predictions for the five free parameters of our orbital evolution model: Callisto's $t_\alpha$, $d$, and $k_2/Q$ and Ganymede's $d$ and $k_2/Q$. A benefit of this work is that we have isolated key, measurable parameters that can be observed by future space missions such as JUICE  to test these predictions. Astrometry measurements would be able to track Callisto's and Ganymede's migration rate, $\dot{a}$, and give us $t_\alpha$. Geodetic measurements can in principle provide the obliquities and tidal responses ($k_2$ and $k_2/Q$) of both satellites, and a combination of induction sounding and tidal measurements place constraints on $d$.

As our discussion in Sections~5.1 and 5.2 illustrates, the thermal evolution and consequences for surface geology are more complicated than we can address here. A particular drawback of our approach is that we assume a constant solid-body $k_2/Q$, while in reality this quantity is expected to change as the thermal state of the satellite changes. It would be of great interest to investigate coupled thermal-orbital evolution scenarios in the context of the resonance locking model.

A drawback of our model is that we use analytical solutions rather than N-body simulations in tracking the orbital dynamics. While this is appropriate for a first look, it may lead to important physics being missed. A recent study has shown that the 2:1 resonance between Ganymede and Callisto could have chaotic effects and pump the eccentricities up as high as 0.1 \citep{lari2020long}. Future work should include N-body simulations to capture details in the orbital dynamics such as these.

Two aspects of such scenarios are of particular interest. One is that Callisto, like Iapetus, could conceivably have frozen in a shape acquired at an earlier time (faster spin rate): a ``fossil bulge'', like the Earth's Moon. This would have important consequences for interpretation of its shape and gravity. The second is that Callisto might have passed through a so-called Cassini State transition around  2.5~Gyr, which again could have had interesting consequences for its thermal evolution.

\section{Conclusion}
\label{section:conclusion}
In this work we have followed two avenues to reconcile Callisto's present-day inclination with an expected short inclination damping time-scale: (1) its physical properties reduced dissipation; or (2) a recent dynamical event increased its inclination. We have shown that despite the uncertainties in Callisto's bottom drag coefficient, ocean thickness, and ice shell thickness, ocean obliquity tides are still strong enough to damp Callisto's inclination within a few billion years, which is incompatible with a primordial inclination. Incorporating a new tidal theory for dissipation in the giant planets, we have found scenarios in which Callisto's inclination and eccentricity are excited by crossing resonances with Ganymede and decay to their present-day values. 
Future measurements of Callisto's semi-major axis migration rate, obliquity, tidal response, and gravity moments will provide stringent tests of this proposed evolution model.

\section*{Acknowledgements}
We thank Benoit Noyelles for the insightful and helpful review. Bruce Bills provided additional comments and suggestions. IM was financially supported by NASA under grant no. 80NSSC20K0570 issued through the NASA Solar System Workings program. This material is based upon work supported by the National Science Foundation Graduate Research Fellowship under Grant No. DGE-1842400.

\section*{Data Availability}

The data underlying this article will be shared on reasonable request to the corresponding author.



\bibliographystyle{mnras}
\bibliography{diss}


\bsp	
\label{lastpage}
\end{document}